# Effects of Mischmetal Composition and Cooling Rates on the Microstructure and Mechanical Properties of Al-(Ce, La, Nd) Eutectic Alloys

Jie Qi[1], Erin C. Bryan[1], David C. Dunand[1]

[1]Department of Materials Science & Engineering, Northwestern University, Evanston, IL, 60208, USA

**Abstract:**

This study investigates the substitution of cerium (Ce) with mischmetal (MM) in cast Al-MM alloys, focusing on microstructure, hardness, tensile/compression properties, creep resistance, and coarsening resistance. Al-MM alloys with various MM compositions (Ce, Ce-50La, Ce-33La, and Ce-27La-19Nd, wt%) exhibit near-eutectic and hyper-eutectic microstructures for Al-9MM and Al-12MM compositions, respectively, with similar as-cast hardness (~525 MPa). All Al-9MM alloys show tensile yield stress ~55 MPa, ultimate tensile strength ~130 MPa, and fracture strain ~8%. The microstructural and mechanical properties consistency demonstrates the flexibility of MM compositions in Al-MM alloys. Al-9MM exhibits excellent coarsening resistance, with minimal hardness reduction when exposed to 300 and 350 °C for up to 11 weeks, and a modest ~15% hardness reduction at 400 °C for 8 weeks, outperforming eutectic Al-12.6Si and Al-6.4Ni alloys. Additionally, Al-9MM shows higher creep resistance at 300 ºC compared to most precipitate-strengthened Al-Sc-Zr and solid-solution-strengthened Al-Mg/Mn alloys, but is outperformed by eutectic-strengthened Al-6.4Ni and Al-10Ce-5Ni alloys. The effect of casting cooling rate is investigated through wedge casting: Al-9Ce transitions from hypo- to hyper-eutectic as cooling rates decrease, while Al-12Ce consistently shows hyper-eutectic microstructures. $Al_{11}Ce_3$ lamellae become finer and more closely spaced with increasing cooling rates. Al-9Ce maintains steady hardness at high to moderate cooling rates but shows reduced hardness at lower rates, whereas Al-12Ce shows no change in hardness. With a 15% reduction in energy consumption and $CO_2$ emissions, Al-Ce alloys where Ce is replaced with MM offer comparable mechanical properties and enhanced environmental benefits, highlighting MM's potential as a sustainable alternative.

## 1. Introduction

Aluminum alloys optimized for high-temperature functional environments are pivotal in aerospace applications due to their exceptional creep- and coarsening resistance, as well as low density and reduced material/operation cost compared to other metallic materials such as titanium alloys. The design of creep-resistant Al alloys involves various strengthening strategies such as: (i) alloying with Mg and Mn for solid-solution strengthening [1–5], (ii) introducing aging-induced nano- or submicron precipitates like $L1_2$-$Al_3(Sc, Zr)$, θ′-$Al_2Cu$, and V-$Al_8Cu_4Sc_1$ [2,6–11], and (iii) forming eutectic phases (e.g., Al-Si, Al-$Al_3Ni$) for load transfer from a weak Al matrix to a stiff eutectic phase skeleton [2,4,8,12–16]. The coarsening resistance stems largely from the reduced solubility and diffusion rates of solute atoms in the Al matrix with phases containing light RE elements (La, Ce, Pr, Nd) and refractory metals (W, Mo, Ta) showing excellent thermal stability [17,18]. Al-Ce-based eutectic compositions are particularly notable for their exceptional resistance to creep and coarsening as well as for good castability [19], with ~11 vol% of "Chinese script-like", lamellar $Al_{11}Ce_3$ phase [20] formed within the Al matrix, strengthening the alloy via load transfer and Orowan strengthening [19].

In addition to the quest for better thermo-mechanical properties, current research increasingly focuses on sustainability in the production of Al-Ce-based alloys. Efforts to reduce energy consumption, environmental burden, and costs have led to exploring alternatives such as substituting Ce with mischmetal, a more eco-friendly and energy-efficient option [21]. Mischmetal is a mixture of light rare-earth elements (LREEs), predominantly Ce and La with smaller quantities of Nd and Pr, which coexist within bastnaesite and monazite ores [22]. During element separation, Nd and Pr, being more economically valuable, are extracted first, resulting in a Ce-La-rich mischmetal of lower economic value. Foregoing the further separation of Ce and La in mischmetal production leads to energy savings and $CO_2$ footprint reduction [22], alongside a decrease in cost. Additionally, mischmetal can be extracted from recycled Ce-La(-Nd) mixtures [23–25], further curtailing costs and environmental impact. Recent developments have also unveiled an intriguing approach in the separation of LREEs such as La, employing microbial methods: this biological extraction technique is very environmentally friendly, marking a significant advancement in the sustainable production of LREEs [26]. Notably, mischmetal composition may vary due to differing source materials.

Replacing Ce with other LREEs in Al-Ce alloys appears feasible based purely on examination of the binary phase diagrams of Al-Ce, La, Nd, and Pr, with the formation of Al-$Al_{11}RE_3$ eutectic structures at similar temperatures and compositions on the aluminum-rich side [27]. Furthermore, stoichiometric $Al_{11}RE_3$ compounds with varying mischmetal compositions in the RE atomic sites maintain consistent crystal structures and closely matched lattice parameters and mechanical properties [20,27]. This mutual solubility among LREEs indicates that using a range of MM compositions in Al-MM eutectic alloys can consistently yield a single $Al_{11}MM_3$ strengthening phase with similar mechanical properties and coarsening resistance, regardless of the precise mischmetal compositions [27]. Currently, only limited experimental support for this hypothesis exists, by the observed metallographic and tensile properties similarities between Al-12Ce and Al-12MM (wt%) alloys [21]. In the present study, we extensively investigate the effects of utilizing different LREE mischmetal compositions in Al-MM eutectic alloys.

Beyond sustainability goals, the effects of casting cooling rates on microstructure and mechanical properties in Al-Ce alloys remain to be fully elucidated with practical significance. The selection of solidification microstructure, described by the Solidification Microstructure Selection Map (SMSM), and the formation of the eutectic structure (the "coupled zone") are influenced by both the composition and the phase growth rates [28–31]. A notable study on Al-14Ce alloys demonstrated that cooling rates below 1600 K/s promote the growth of the primary $Al_{11}Ce_3$ phase [30]. However, further investigations are required to quantitatively determine how cooling rates affect the microstructure of Al-Ce alloys, specifically the thickness and spacing of $Al_{11}Ce_3$ lamellae. Additionally, the impact of these microstructural variations on the mechanical properties of the alloys needs thorough exploration.

In this article, we focus on two primary areas of investigation about Al-RE alloys. First, we investigate the effect of various mischmetal compositions on the eutectic composition, microstructure, coarsening resistance, and mechanical properties of Al-RE alloys; three binary compositions are examined for RE: Ce-50La and Ce-33La representing typical composition ranges of Nd- and Pr-depleted mischmetal; and Ce-27La-19Nd (denoted as MM) representing an average mischmetal composition from natural resources. [27] Second, we wedge-cast Al-9Ce and Al-12Ce alloys to achieve a range of cooling rates, and then examine the relationships among alloy composition, cooling rates, microstructure, and mechanical properties.

## 2. Experiment Methods

Al-RE(Ce, La, Nd) alloys were cast by melting and mixing different master alloys: Al-8Ce, Al-15Ce, Al-15La, and Al-15Nd (all compositions are thereafter given in wt%). The Al-8Ce and Al-15Ce master alloys were procured from Eck Industries (Manitowoc, WI). We arc-melted the Al-15La, and Al-15Nd master alloys with pure elements: La and Nd (both > 99.9% purity, from Luciteria Science, Olympia, WA), and high-purity Al (99.99%, from C-KOE Metals, Euless, TX). Arc-melting was performed under an argon atmosphere using water-cooled copper crucibles. Each 30 g ingot of the master alloy was melted with a 400A current for 30 s. This melting process was repeated five times, flipping the ingot after each cycle to promote uniformity.

The melting of the final alloy was conducted in a graphite crucible kept at 750 ºC in ambient air condition. The mixture of mater alloys was melted, stirred, and de-slagged regularly for 1 h to ensure complete homogenization. Subsequently, the de-slagged melt was poured into a graphite mold preheated to 200 ºC and placed on an ice-cooled copper plate to achieve directional solidification, resulting in four cylindrical ingots, each measuring 15 mm in diameter and 60 mm in length. Samples subjected to thermal exposure were kept at temperatures of 300, 350, and 400 ºC for up to 11 weeks, followed by water quenching. For the wedge casting process, 1 kg of the alloy was prepared by combining and melting pure

Al and Al-15Ce master alloy in a graphite crucible at 800 ºC, under ambient air conditions. The alloy was stirred and de-slagged periodically over 2 h to ensure homogeneity. The molten alloy was then cast into a low-carbon-steel wedge mold which was preheated to 200 ºC and subsequently coated with boron nitride for ease of demolding. Cooling rates along the centerline of the wedge mold were measured by K-type thermocouples with data recorded by a DATAQ Di-245 logger at 25 Hz. All testing specimens were extracted along the centerline of the wedge ingot, thereby avoiding the regions with higher cooling rates near the surfaces of the mold.

Samples for Scanning Electron Microscope (SEM) analysis, etching, and hardness measurements were polished to a 1 μm surface finish, followed by vibrational polishing with 0.06 μm colloidal silica. A deep etching solution with 20 wt% NaOH and 80 wt% distilled water was utilized. The microstructural and composition measurements were performed using Quanta 650 and JEOL JSM-7900FLV SEMs, both equipped with Energy-Dispersive X-ray Spectroscopy (EDS) detector. Electron Backscattered Diffraction (EBSD) analysis was carried out on the Quanta 650 SEM. The measurement of Vickers microhardness was on a Buehler Wilson VH3100 hardness tester, with a consistent load of 200 g maintained for 5 s. Hardness values were measured at twenty points per sample, with average values and standard deviations reported. Tensile tests were carried out at Westmoreland Mechanical Testing & Research, Inc. (Youngstown, PA), following ASTM E8/E8M-22 standards, with strain rates of $10^{-5}$ and $10^{-4}$ s$^{-1}$ pre- and post-yield, respectively.

As-cast samples were machined using a lathe to creep specimens with diameters of ~5 mm and heights of ~10 mm. Each specimen was placed between boron-nitride-coated tungsten carbide plates within a compression cage within a furnace. Temperature was monitored by a thermocouple touching the specimen surface. The compressive stress was applied through direct tensile loading of the compression cage after calibration with a load cell. A linear variable displacement transducer (LVDT) was used to measure the compressive deformation of the sample. Upon attaining a steady-state strain rate, the stress was increased to the next level. This process continued until the total strain reached ~10%, at which point the creep test was terminated.

The 2D axisymmetric finite element model (FEM) was constructed using the COMSOL Multiphysics software™ to calculate the cooling rates at different heights along the centerline of the wedge sample. The FEM component uses actual sample dimensions, with parameters included in Table S1 (Supplementary Information).

The Al$_{11}$Ce$_3$ lamellar widths and spacings were quantified on SEM images using an image processing MATLAB script. For each sample, approximately 400-600 data points for width or spacing were recorded. Subsequently, the mean and standard deviation of these measurements were calculated to provide a comprehensive statistical analysis.

For compression testing at ambient temperature, cylinders were electrical discharge machined (EDM) from wedge samples, with dimensions of 5 mm in diameter and 10 mm in height. During compression, samples were positioned between two boron-nitride-coated tungsten carbide plates in an MTS 5 servo-hydraulic compression testing machine. Tests were carried out to 10% compressive strain at a strain rate of $10^{-4}$ s$^{-1}$. Compressive yield stress was calculated with the 0.2% strain offset method.

## 3. Results and Discussion

### 3.1. Al-Mischmetal Alloys
#### 3.1.1. Microstructures

Thermo-Calc (*TCAL8: Al-Alloys V8.1 database*) was used to calculate the initial eutectic compositions for various Al-RE systems with different mischmetal compositions in RE sites, as illustrated in the phase diagrams in Figure S1. Here, Al-9RE and Al-12RE are predicted to be hypo- and near-eutectic compositions, respectively. To explore the impact of the Ce/La ratio on eutectic composition, microstructure, and mechanical properties, Nd- and Pr-free mischmetal compositions CeLa and 2CeLa are employed at first, resulting in alloy compositions Al-4.5Ce-4.5La, Al-6Ce-6La, Al-6Ce-3La, and Al-8Ce-4La, spanning from Al-9RE to Al-12RE. The microstructures are depicted in Figure 1(a-d). The two Al-9RE alloys in Figure 1(a,c) exhibit small regions of primary Al phase (~8 vol% determined by Image J, indicated by yellow arrows) in the left images. The primary Al regions are evenly distributed within the ingot cross section. The high-magnification micrographs on the right illustrate the "script-like" eutectic structure. The two Al-12RE alloys shown in Figure 1(b,d) are hyper-eutectic with primary $Al_{11}RE_3$ phases outlined in red boxes. No notable difference in microstructure is apparent when varying Ce/La ratios. Al-9MM was further cast to investigate the effects of Nd additions, with MM denoting the average mischmetal composition Ce-27La-19Nd. Al-9Ce was also cast as a control alloy. Similar to the other Al-9RE alloys (Figure 1(a,c)), Al-9MM and Al-9Ce display slightly hypo-eutectic microstructures (Figure 1(e,f)). Notably, Ce, La, and Nd have mutual solubility in the $Al_{11}RE_3$ phase (as proved by the EDS map scan in Figure 1i), contradicting the Thermo-Calc-predicted phase separation of $Al_{11}RE_3$ into $Al_{11}Ce_3$ and $Al_{11}La_3$ (Figure S1), and aligning with the conclusions from a previous study focused on pure $Al_{11}RE_3$ compounds [27]. Al-4.5Ce-4.5La and Al-6Ce-6La underwent further deep etching to remove the Al matrix and highlight the $Al_{11}RE_3$ phase. In the hypo-eutectic Al-4.5Ce-4.5La (Figure 1g), $Al_{11}RE_3$ lamellae are aligned within eutectic colonies whose boundary is highlighted by a yellow dashed line. In hyper-eutectic Al-6Ce-6La (Figure 1h), apart from the eutectic $Al_{11}RE_3$ phase, faceted primary $Al_{11}RE_3$ phases are highlighted with red arrows.

The primary $Al_{11}RE_3$ phases in Al-12RE alloys are always surrounded by Al dendrites (Figure 1(b,d)), consistent with the early-solidified primary $Al_{11}RE_3$ particles depleting RE elements in the surrounding liquid and acting as the heterogeneous nucleation site of primary α-Al dendrites. The large size and spacing of primary $Al_{11}RE_3$ phases reduce the Orowan strengthening effect, and their sharp edges increase stress concentration and facilitate microcrack formation, thereby decreasing ductility. Furthermore, the volume fraction of the more strengthening-effective lamellar $Al_{11}RE_3$ in eutectic regions does not increase with RE additions beyond the eutectic composition. EDS analysis confirms that the eutectic regions in all four alloys approximate an Al-10RE composition, with the surplus RE in Al-12RE alloys precipitating into primary $Al_{11}RE_3$ phases. The efficacy of load transfer during mechanical loading largely depends on the volume fraction of $Al_{11}RE_3$ in the eutectic regions. Consequently, Al-12RE alloys may not exhibit evident improvement in load transfer compared to Al-9RE alloys, as evidenced by the similar hardness of Al-12RE and Al-9RE (detailed in the later Section 3.1.2 and Figure 2a).

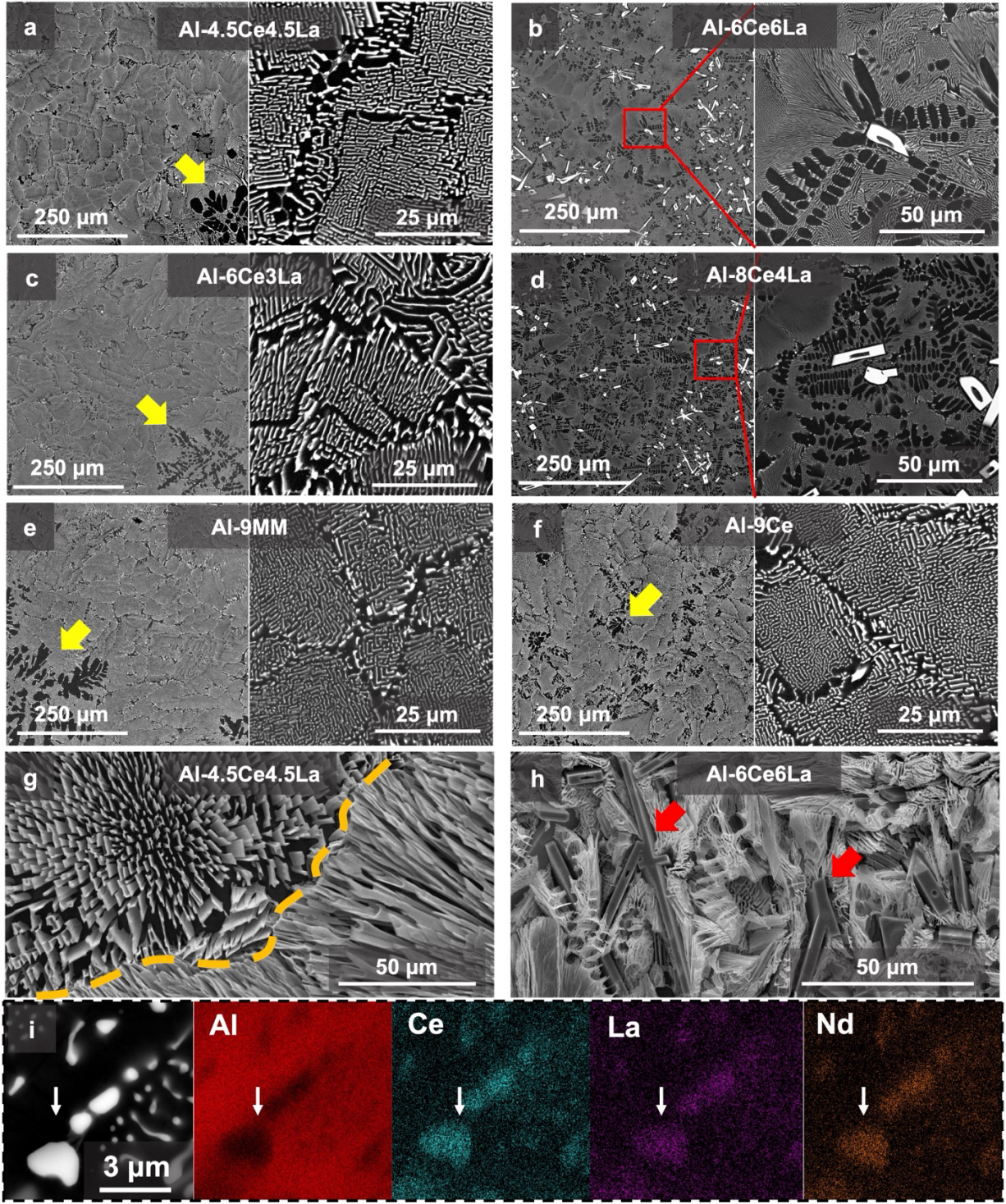

Figure 1. SEM images of microstructures for Al-4.5Ce-4.5La, Al-6Ce-6La, Al-6Ce-3La, Al-8Ce-4La, Al-9MM, and Al-9Ce. (a,c,e,f) slightly hypo-eutectic Al-9RE alloys, with yellow arrows pointing to primary Al dendritic regions (black contrast). (b,d) hyper-eutectic Al-12RE alloys, with magnified views of primary $Al_{11}RE_3$ phases shown in red boxes. (g) deep-etched hypo-eutectic Al-4.5Ce-4.5La, showing eutectic lamellar $Al_{11}RE_3$ phase, with the yellow dashed line highlighting the eutectic colony boundary. (h) deep-

etched hyper-eutectic Al-6Ce-6La, with red arrows pointing to primary $Al_{11}RE_3$ phases embedded within eutectic regions (with eutectic lamellar, interconnected $Al_{11}RE_3$ phase). (i) EDS map scan of $Al_{11}MM_3$ phase in of Al-9MM (white arrow) showing mutual solubility of Ce, La, and Nd.

Table 1. EDS-determined compositions (in wt%) of alloys studied in Figure 1.

| Nominal Composition* | Ce | La | Nd | Ce+La+Nd |
|---|---|---|---|---|
| Al-4.5Ce-4.5La | $4.60 \pm 0.14$ | $4.22 \pm 0.13$ | - | $8.82 \pm 0.19$ |
| Al-6Ce-3La | $6.42 \pm 0.18$ | $2.91 \pm 0.08$ | - | $9.33 \pm 0.20$ |
| Al-6Ce-6La | $5.54 \pm 0.19$ | $5.59 \pm 0.10$ | - | $11.13 \pm 0.21$ |
| Al-8Ce-4La | $7.84 \pm 0.13$ | $3.79 \pm 0.06$ | - | $11.63 \pm 0.14$ |
| Al-9Ce | $9.50 \pm 0.14$ | - | - | $9.50 \pm 0.14$ |
| Al-9MM | $5.20 \pm 0.10$ | $2.35 \pm 0.07$ | $1.70 \pm 0.28$ | $9.25 \pm 0.31$ |

* Al balance

### 3.1.2. Mechanical Properties and Hardness Retention under Thermal Exposure

The hardness values of as-cast Al-4.5Ce-4.5La, Al-6Ce-3La, Al-6Ce-6La, Al-8Ce-4La, Al-9MM, and Al-9Ce alloys are ~525 MPa, as shown in Figure 2a, with differences within experimental errors. The Al-12RE alloy does not show a pronounced increase in strength compared to the Al-9RE alloy, as discussed in Section 3.1.1. Tensile tests of the Al-9RE alloys - specifically Al-4.5Ce-4.5La, Al-6Ce-3La, and Al-9MM - are presented in Figure 2c. Al-9RE alloys show an average yield stress of ~55 MPa, ultimate tensile strength (UTS) of ~130 MPa, and fracture strain close to 8 %. Different mischmetal compositions do not appear to affect the hardness, tensile properties, and microstructure of Al-RE alloys. This compositional adaptability may eliminate the need for adjusting the mischmetal composition from various sources.

High-temperature thermal exposure tests were conducted to assess the coarsening resistance of the alloys. As shown in Figure 2a, after exposure to 400 ºC (~73% of their absolute eutectic temperature) for up to 8 weeks, the hardness of the Al-RE alloys decreases by 15%, from ~525 to ~450 MPa. In contrast, eutectic Al-6.4Ni [32] and Al-12.6Si [19] alloys experience greater hardness reductions (27 and 40%, respectively), albeit from higher as-cast hardness values. All alloys, including Al-6.4Ni, Al-12.6Si, and Al-RE, exhibit similar hardness after the 8-week exposure. Additionally, Al-RE and Al-6.4Ni show a steadier hardness reduction in the time-log scale, while Al-12.6Si exhibits more rapid initial softening (~30% reduction in 2 days). Given the similar hardness evolution trends of alloys across different RE compositions, the thermal exposure tests at 300 and 350 ºC were only conducted for Al-9MM and the control Al-9Ce alloy. As shown in Figure 2b, the alloys show marginal hardness reduction, remaining within the range of experimental errors, for both temperatures. This study demonstrates that the Al-RE alloys exhibit excellent strength retention and corresponding coarsening resistance up to 400 ºC, well above typical operational temperatures for current Al-based alloys.

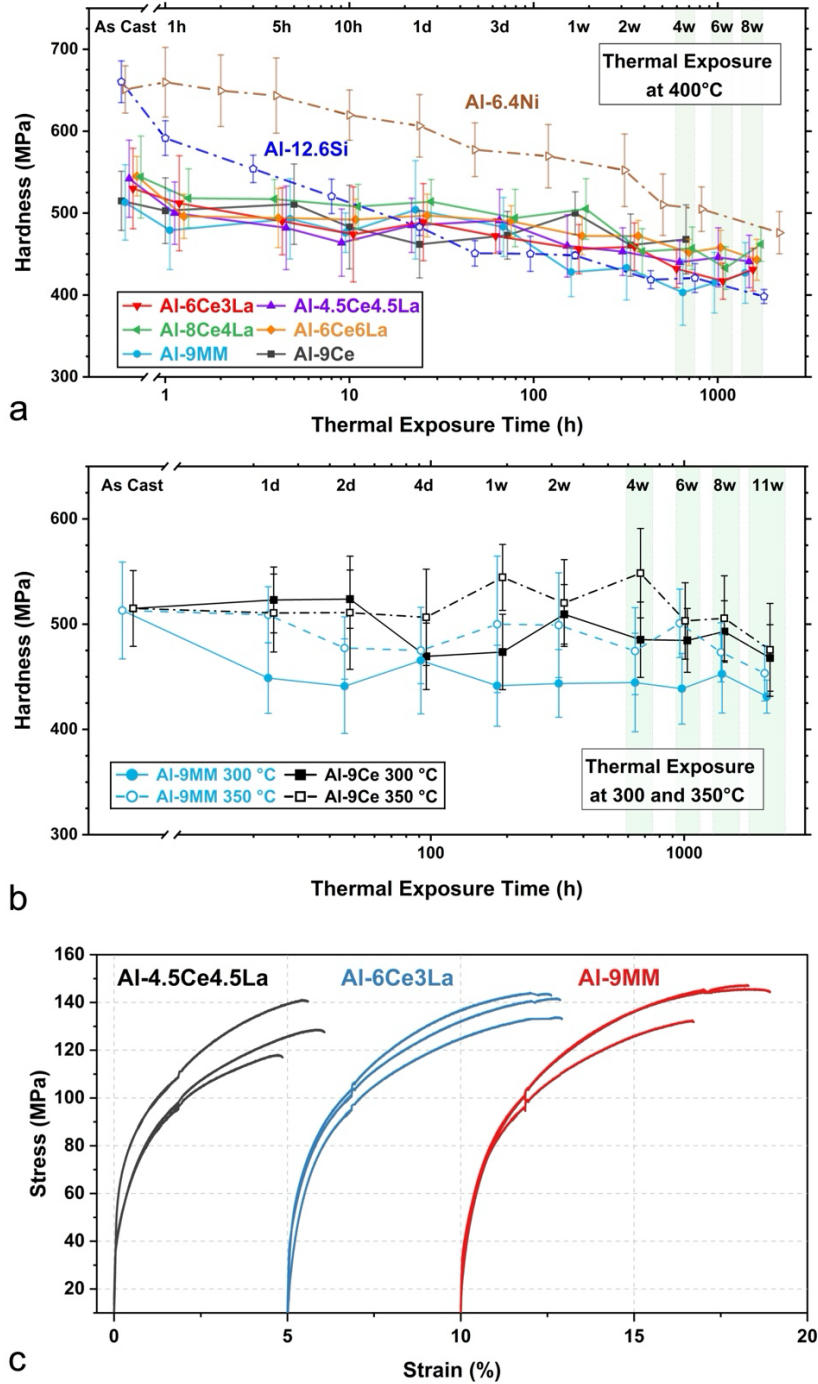

Figure 2. (a) Hardness evolution after 400°C thermal exposure for up to 8 weeks for Al-4.5Ce4.5La, Al-6Ce-3La, Al-6Ce-6La, Al-6Ce-6La, Al-9MM, and Al-9Ce. Literature data for typical eutectic systems Al-6.4Ni [32] and Al-12.6Si [19] are plotted for comparison. (b) Hardness evolution of Al-9MM and Al-9Ce alloys after thermal exposure for up to 11 weeks at 300 and 350 °C. For clarity, data points from different alloys corresponding to the same duration of thermal exposure are slightly offset horizontally in (a,b). (c) Tensile stress-strain curves for as-cast Al-4.5Ce-4.5La, Al-6Ce-3La, and Al-9MM (3 samples per composition). A small discontinuity at 2% strain is due to a jump in strain rate.

The negligible hardness reduction in Al-MM following exposure at 300 and 350 ºC is attributed to the excellent coarsening resistance of the $Al_{11}MM_3$ phase. Figure 4 compares the microstructures of Al-9MM in various conditions: as-cast, 11-week exposure at 300 and 350 ºC, and 8-week exposure at 400 ºC. Figure 4(b,c) show minimal $Al_{11}MM_3$ phase coarsening at 300 and 350 ºC. The coarsening of the $Al_{11}MM_3$ phase can be approximated by the Ostwald ripening equation [33]:

$$\Delta(r^3) = \frac{A\,D_{(T)}\,C_{(T)}}{R\,T}\,t \qquad [1]$$

where r is the precipitate size; A is a constant containing the Al–$Al_{11}MM_3$ interfacial energy and $Al_{11}MM_3$ phase volume fraction; $D_{(T)}$ and $C_{(T)}$ are the diffusivity and solubility of LREEs in Al matrix at temperature T; R is the gas constant; T is the temperature; and t represents the thermal exposure time. The comparison of $D_{(T)}$ and $C_{(T)}$ for LREEs, Ni, and Si is provided in Figure 3. Notably, the $D_{(T)}$ and $C_{(T)}$ for LREEs are considerably lower than those for Ni and Si, which accounts for the enhanced coarsening resistance of Al-MM alloys compared to Al-Ni and Al-Si eutectic alloys, as illustrated in Figure 2a. Furthermore, at the elevated temperature of 400 ºC, the increase in the $D_{(T)}×C_{(T)}$ term in Eq.[1 leads to more pronounced coarsening. This is evidenced by the spheroidization of precipitates and their morphological transition from lamellar to short rod-like structures, as shown in Figure 4d. This morphological transition is driven by the reduction of Al–$Al_{11}MM_3$ interface area and associated interfacial energy, as a short rod-like morphology is more stable than a lamellar structure when the volume fraction of precipitates falls below the threshold of $1/\pi$ (~32%) [34]. The Al-9MM alloy, with an estimated 11 vol% of the strengthening phase, exemplifies this morphological preference.

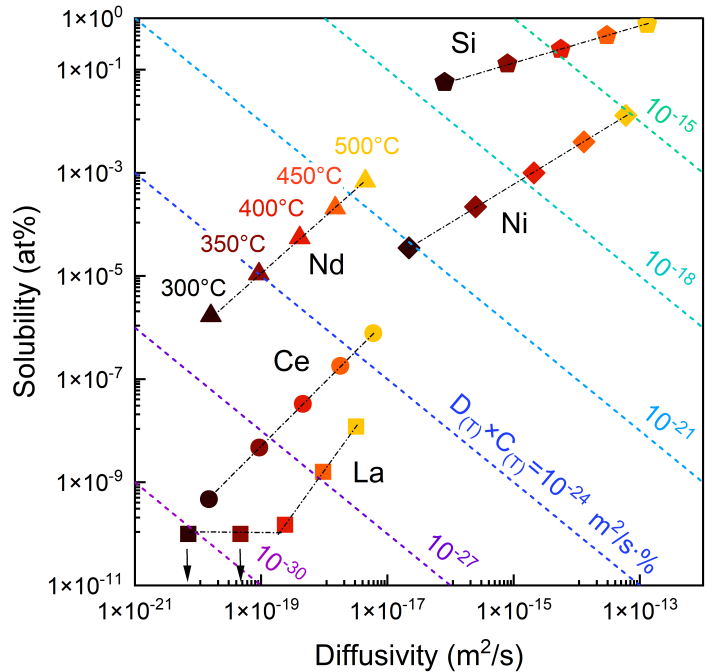

Figure 3. Plot of diffusivity vs. solubility of LREEs (La, Ce, Nd), Ni, and Si in Al at various temperatures between 300 and 500 ºC. Symbol colors represent different temperatures, with the values labeled for the Nd series. Contour lines of constant values of the product of diffusivity and solubility ($D_{(T)}×C_{(T)}$) are also plotted. Diffusivities $D_{(T)}$ are calculated based on the Arrhenius equation with data from Ref. [17,35] and solubilities $C_{(T)}$ are derived from binary phase diagrams generated *via* Thermo-Calc (version 2022b) with TCAL8 database.

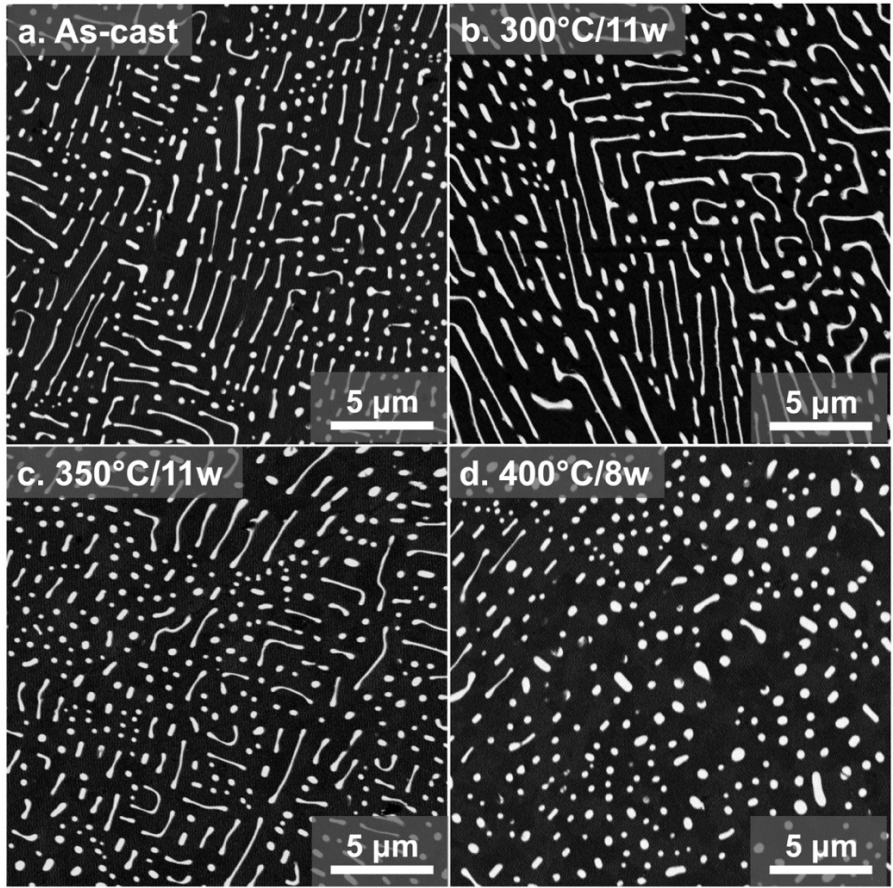

Figure 4. Comparison of Al-9MM microstructure in (a) as-cast state, and after thermal exposure of (b) 300 °C for 11 weeks, (c) 350 °C for 11 weeks, and (d) 400 °C for 8 weeks. The Al matrix and $Al_{11}MM_3$ phases are in dark and bright contrasts respectively.

### 3.1.3. Creep Resistance

The creep behavior of Al-9MM alloys at 300, 350, and 400 °C is shown in Figure 5a. An illustrative creep curve showing strain vs. time, with primary and secondary creep regimes for the minimum creep strain rate determination, is available in Figure S3a. The minimum creep strain rate ($\dot{\varepsilon}$) is plotted against the applied stress ($\sigma$) in a double-logarithmic plot consistent with the Norton creep power law:

$$\dot{\varepsilon} = A\ \sigma^n \exp(-\frac{Q}{RT}) \qquad [2]$$

where A is a dimensionless constant, n is the apparent stress exponent, Q is the creep activation energy, R is the gas constant, and T is the temperature. The data points in Figure 5a at the lowest measured stresses appear to belong to a transition regime between high-stress dislocation creep and low-stress diffusional creep, as reported in other studies [19,36].

For the high-stress regime controlled by dislocation creep, the apparent stress exponent is calculated as n=9 for Al-9MM. The apparent stress exponent is greater than $n_{Al} = 4.4$ [37], the dislocation stress exponent for pure Al. This indicates the presence of a dislocation threshold creep stress ($\sigma_{th}$) below which significant time-dependent dislocation creep deformation is inhibited. The Norton law is then modified to:

$$\dot{\varepsilon} = A' \, (\sigma - \sigma_{th})^{n_{Al}} \exp\left(-\frac{Q_{Al}}{RT}\right) \tag{3}$$

with A' as a dimensionless constant, and $Q_{Al}$ as the aluminum-specific activation energy. The $\sigma_{th}$ values are calculated as 18, 14, and 12 MPa for 300, 350, and 400 ºC, respectively, by using the modified Norton Law (as illustrated in Figure S3b). The low-stress diffusional-creep points are excluded from the calculation of $\sigma_{th}$ and from the fitting of best-fit lines for determining the apparent stress exponent. The dislocation creep curves fitted by the modified Norton Law (Eq. [3]) are plotted as solid lines in Figure 5a.

For the diffusional creep region at low applied stresses, the strain rate equation, by summing the lattice and grain boundary diffusion, can be modeled by Coble and Nabarro-Herring equations [38]:

$$\dot{\varepsilon} = \frac{14\sigma\Omega}{kTd^2} D_v \left[1 + \frac{\pi\delta}{d}\frac{D_b}{D_v}\right] \tag{4}$$

where $\Omega$ is the atomic volume of Al, $k$ the Boltzmann constant, d the grain size, $D_v$ the lattice self-diffusivity of Al, $\delta$ the effective thickness of the grain boundary, and $D_b$ the grain boundary diffusion coefficient. These parameters, sourced from Ref. [38], along with the EBSD-determined grain size d = 250 ± 100 μm (Figure S2), allow for the calculation of the diffusional creep rate, which is plotted as bands in Figure 5a.

The top figure of Figure 5b shows an Arrhenius plot of $\log(\dot{\varepsilon})$ vs. $\frac{1}{T}$ at applied stresses of 20 and 23 MPa. According to Eq. [2, the apparent activation energy Q is calculated as 220 ± 8 kJ/mol from these two lines. The activation energy of Al-9MM is comparable to that of the eutectic Al-12.5Ce alloy [19], but much higher than that of pure Al (151 kJ/mol) [39], which is also indicative of a threshold stress. Similarly, the bottom plot in Figure 5b is based on the modified Norton law (Eq. [3]) with ($\sigma$-$\sigma_{th}$) values of 8 and 11 MPa. The activation energy, $Q_{Al}$, is then calculated as 156 ± 5 kJ/mol, consistent with the literature value of 151 kJ/mol for pure Al [39]. Figure 5c shows the double logarithmic plot of $\dot{\varepsilon}/\exp\left(-\frac{Q_{Al}}{RT}\right)$ vs. ($\sigma$-$\sigma_{th}$) for Al-9MM alloys at 300, 350, and 400 ºC. According to Eq. [3, the high-stress creep data for dislocation creep at various temperatures will ideally align along a single line, reflecting the consistent apparent stress exponent $n_{Al}$ = 4.4, activation energy $Q_{Al}$=156 kJ/mol, and pre-exponential factor A'. The observed slight broadening of the distribution band is due to the propagation of experimental errors.

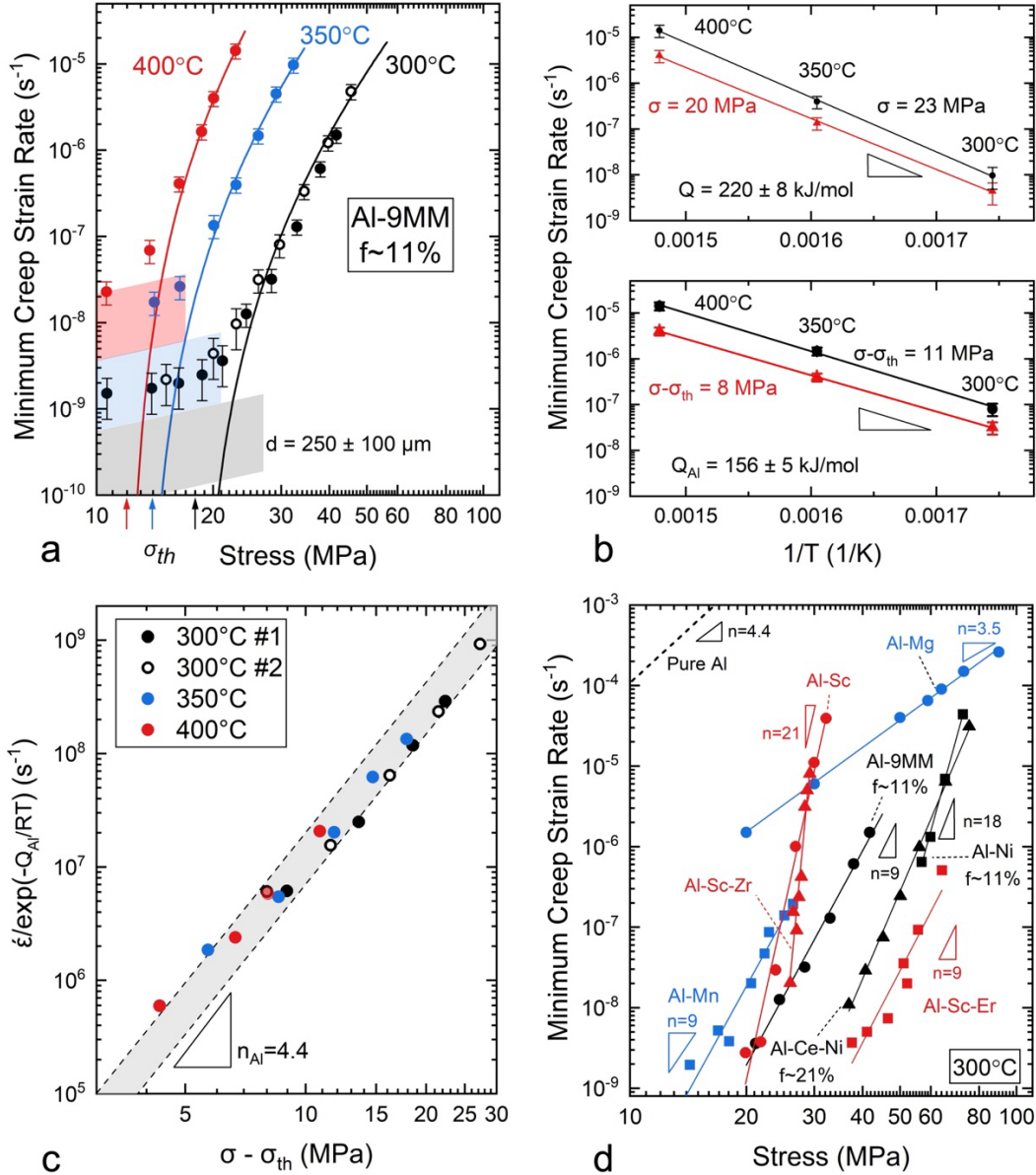

Figure 5. (a) Double logarithmic plot of minimum creep strain rate vs. compressive stress for Al-9MM alloys at 300, 350, and 400°C, Arrows on the x-axis indicate the dislocation threshold creep stress for each temperature. Solid and hollow data points for the 300 °C test correspond to the first and second samples tested. The lines and bands are the fitted dislocation creep (Eq.[3]) and Coble and Nabarro-Herring diffusional creep (Eq.[4]). (b) Arrhenius plot of minimum creep strain rate vs. reciprocal temperature at σ = 20 and 23 MPa (top figure) and (σ-σ_th) = 8 and 11 MPa (bottom figure). The curves exhibit consistent slopes, from which the creep activation energies Q for Al-9MM (top figure) and $Q_{Al}$ for Al (bottom figure) are determined by Eq.[2 and 3, respectively. (c) Double logarithmic plot of $\dot{\varepsilon}/\exp\left(-\frac{Q_{Al}}{RT}\right)$ vs. (σ-σ_th) for Al-9MM alloys at 300, 350, and 400°C. According to Eq.[3], the high-stress creep data for dislocation creep under different temperatures fall along a narrow band with stress exponent $n_{Al}$ = 4.4. (d) Comparative analysis of the creep resistance of Al-9MM with Al-5Mg [40], Al-1Mn [41], Al-10Ce-5Ni [16], Al-6.4Ni [32], Al-0.3Sc [42], Al-0.24Sc-0.04Zr [42], and Al-0.11Sc-0.14Er [43]. The low-stress diffusional creep regime for Al-9MM, shown in (a), is omitted for clarity. Apparent stress exponents *n*, as the slopes of best-

fit lines (Eq.[2], are indicated. In (a) and (b), the total volume fractions (f) of the eutectic phases $Al_{11}MM_3$ and $Al_3Ni$ (when Ni is present) are labeled.

The high apparent stress exponent n=9 of Al-9MM indicates that precipitate-dislocation interactions primarily limit the creep rate [38]. Two active mechanisms, precipitation strengthening and load transfer, control the creep rate in Al-9MM eutectic-strengthened alloys. Precipitation strengthening relies on inhibiting dislocation climb over precipitates. This mechanism is also present in $L1_2$ precipitate-strengthened Al alloys, such as the Al-0.3Sc [42], Al-0.24Sc-0.04Zr [42], and Al-0.11Sc-0.14Er [43] (wt%), plotted in Figure 5d. Despite the excellent creep resistance of Al-0.11Sc-0.14Er, attributed to the larger precipitate-matrix lattice mismatch from Er alloying [43], most $L1_2$ precipitate-strengthened Al alloys exhibit lower creep resistance than Al-9MM, potentially due to the following reasons. First, the $Al_{11}MM_3$ skeleton in Al-9MM effectively carries the load transferred from the weak Al matrix during creep, reducing the stress-driving dislocation movement in the Al matrix [44]. Second, the high aspect ratio of $Al_{11}MM_3$ platelet strongly inhibits dislocation climb [45], making dislocation bowing the most likely active strengthening mechanism $via$ Orowan strengthening. Third, unlike the Al-$L1_2$ precipitates' coherent matrix-precipitate interface, the incoherence of the Al-$Al_{11}MM_3$ interface may lead to a detachment stress [46] and high strain field in the Al matrix near the interface [19], both hindering dislocations detaching and gliding away from precipitates after bowing. When compared to other eutectic-strengthened Al-based alloys, the volume fraction of the eutectic phase that determines the load transfer ability, and the morphology of the strengthening phase that determines the inhibition of dislocation mobility through Orowan strengthening, become the main factors controlling creep resistance. Examples are the Al-10Ce-5Ni [16] and Al-6.4Ni [32] eutectic alloys with similar creep resistance, featuring ~21 and ~ 11 vol % of fine fibrous strengthening phases respectively, exhibiting much-improved creep resistance as compared to Al-9MM with ~11 vol % of relatively thicker lamellar phase. Finally, a comparison is made in Figure 5d with the solid-solution strengthened alloys, with Mg and Mn as the common alloying elements. The diffusivities of Mn and Mg in Al are $9.5\times10^{-22}$ and $1.4\times10^{-16}$ m$^2$/s, respectively, as calculated using the Arrhenius equation [17,35] with parameters from ref.[17,47]. As they diffuse much more slowly, Mn solute atoms provide a stronger viscous-drag for dislocation glide than Mg, resulting in a greatly improved creep resistance, as shown by the Al-5Mg [40] and Al-1Mn [41] alloys in Figure 5d. However, as dislocation gliding through the matrix is typically easier than bypassing precipitates, the solid-solution-strengthened alloys are generally less creep-resistant than the eutectic-strengthened Al-9MM.

The comparison in Figure 5d is limited to alloys with a single strengthening mechanism among precipitate, eutectic, or solid solution strengthening. This comparison highlights that eutectic systems show promise for high-temperature creep properties. Building on this insight, hybrid strengthening mechanisms have been shown to effectively enhance the creep resistance of Al alloys, as exemplified by the highly creep-resistant Al-Ce-(Ni-)Mn-Sc-Zr alloys with three (four) strengthening mechanisms presented [5,48].

## 3.2.    Effects of Cooling Rates on Microstructure and Mechanical Properties of Cast Al-Ce

### 3.2.1.   Microstructure Variation with Cooling Rate

The near-eutectic Al-9Ce and hyper-eutectic Al-12Ce were wedge-cast to achieve a range of cooling rates. Six sections along the vertical centerline of the wedge-cast Al-9Ce and Al-12Ce samples were imaged and analyzed. Their positions and corresponding cooling rates as determined experimentally or estimated from the fitted relationship (details in Figure S4-5 in supplementary notes) based on experimental results, are shown in Table 2. The compositions of each section, as determined via SEM-EDS, are plotted in Figure 6a. Generally, a higher concentration of heavier solute elements tends to accumulate at the bottom of the wedge-casting mold. A similar trend exists in Figure 6a, though the composition variation remains minimal and within experimental error. This compositional uniformity allows the cooling rate to be isolated as the primary factor influencing microstructure and mechanical properties.

Figure 7 illustrates the microstructures of sections 1, 3, and 5, which correspond to the bottom, middle, and top of the wedge, with the high, intermediate, and low cooling rates, respectively. In Figure 7(a-c), Al-9Ce exhibits a transition from hypo- to near- to hyper-eutectic as the cooling rate decreases. In contrast, Al-12Ce displays hyper-eutectic microstructures for all cooling rates, as shown in Figure 7(d-f). The primary $Al_{11}Ce_3$ phase is consistently surrounded by primary α-Al regions, as expected from the early-solidified primary $Al_{11}Ce_3$ particles depleting Ce atoms in the surrounding liquid and acting as heterogeneous nucleation sites for primary α-Al dendrites.

A more in-depth microstructural analysis was performed on $Al_{11}Ce_3$ lamella thickness and spacing, as finer eutectic precipitates with smaller spacing can enhance the Orowan strengthening. To obtain comprehensive statistics, approximately 500 $Al_{11}Ce_3$ lamellae were measured per section for both compositions using a custom MATLAB image processing script. The results are presented in Figure 6(b,c). First, the high standard deviations reflect regional variation or inhomogeneity within each section, with standard deviations increasing as cooling rates decrease. This phenomenon is illustrated in the SEM image of slow-cooled Section 5 for Al-9Ce (Figure 7c), where regions with sparse (coarse) and fine eutectic structures are highlighted. It is hypothesized that the heat released during solidification leads to uneven thermal gradient and solidification rate distribution, resulting in microstructural variation. This phenomenon is more evident in sections with lower cooling rates. Despite this variability, Figure 6(b,c) reveal a general trend: as expected, higher cooling rates result in smaller lamella spacing and thickness. The rapid solidification reduces the diffusion time for Ce atoms which leads to smaller lamella spacing, while the limited growth time for precipitates results in reduced lamella thickness [49].

Table 2. Section index with the corresponding section positions (heights from the tip of the wedge sample) and cooling rates at the liquidus temperature.

| Section Index | 1 | 2 | 3 | 4 | 5 | 6 |
|---|---|---|---|---|---|---|
| Height (mm) | 5 | 40 | 65 | 85 | 100 | 120 |
| Liquidus Cooling Rate (°C /s) | 89* | 22* | 17** | 15** | 14* | 12** |

\* Values measured from experiment.
\*\*Values estimated from the fitted relationship between cooling rates and height (see Figure S4-5 in supplementary notes).

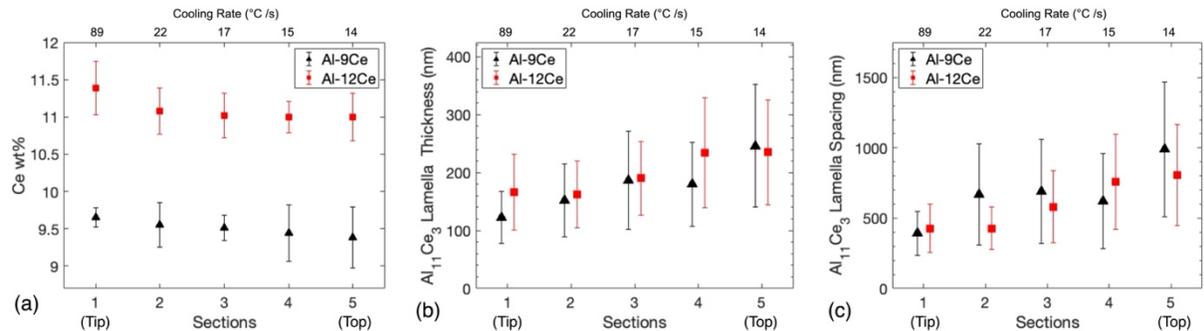

Figure 6. Variation across sections 1 to 5 for wedge-cast Al-9Ce and Al-12Ce alloys of (a) composition, (b) average $Al_{11}Ce_3$ lamellar thickness, and (c) average $Al_{11}Ce_3$ lamellar spacing. Sections 1 and 5 correspond to the tip and top sections of the wedge samples, with the highest and lowest cooling rates, respectively.

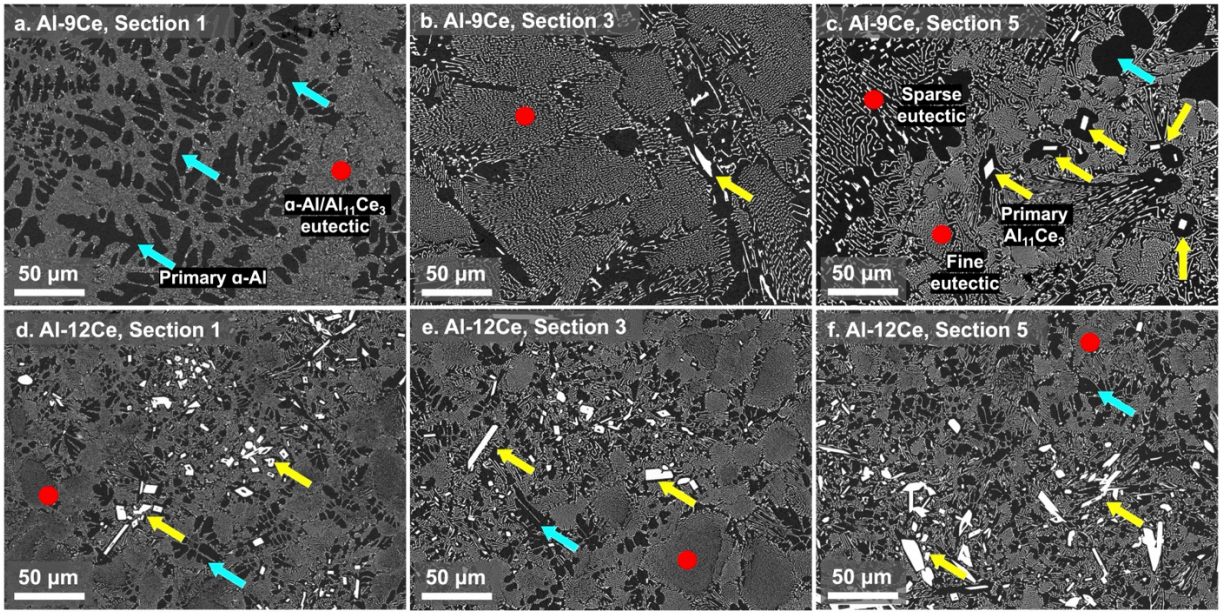

Figure 7. Microstructures of Sections 1, 3, and 5 (from fastest to slowest cooling) in the wedge-cast alloys for (a-c) Al-9Ce; (d-f) Al-12Ce. Examples of primary α-Al dendrites, primary $Al_{11}Ce_3$ phases, and eutectic α-Al/$Al_{11}Ce_3$ regions are highlighted by blue arrows, yellow arrows, and red circles. Two red circles, representing the sparse and fine eutectic regions, are labeled in (c).

### 3.2.2. Mechanical Properties Variation with Cooling Rates

The microhardness evolution of wedge-cast Al-9Ce and Al-12Ce alloys is depicted in Figure 8a. The Al-9Ce alloy demonstrates a ~10% reduction in hardness from sections 1-4 to sections 5-6. As depicted in Figure 7(a,b), sections 1-4 feature a transition from hypo- to near-eutectic microstructures. Although section 1 has a higher volume percentage of weaker primary Al regions, it also exhibits eutectic regions with fine $Al_{11}Ce_3$ lamellae and small spacings (as shown in Figure 6b-c), which are stronger than those in other sections. These two contrasting factors result in relatively consistent hardness values across sections 1-4. In contrast, the hyper-eutectic sections 5-6 predominantly contain primary $Al_{11}Ce_3$ phases, which reduce the volume fraction of the lamellar $Al_{11}Ce_3$ strengthening phase in the entire sample. Additionally, the presence of primary $Al_{11}Ce_3$ phases leads to an increase in weaker primary Al regions surrounding these phases, further diminishing overall hardness. These combined effects result in the observed hardness reduction in these sections. This analysis underscores that the mechanical strength of near-eutectic Al-Ce alloys is predominantly determined by the volume fraction of the lamellar $Al_{11}Ce_3$ strengthening phase.

By contrast, Al-12Ce exhibits consistent hardness values (Figure 8a) across its various sections, which aligns with the similar microstructures shown in Figure 7(d-e). Despite an additional 3 wt% Ce content, Al-12Ce is not stronger than Al-9Ce in sections 1-4. This observation supports the conclusion from Figure 2a, indicating that the additional Ce only forms primary $Al_{11}Ce_3$ phases without increasing the volume fraction of the lamellar $Al_{11}Ce_3$ strengthening phase in the eutectic regions. In sections 5-6, Al-12Ce becomes stronger than Al-9Ce, because the latter has a lower volume fraction of lamellar $Al_{11}Ce_3$ phase in the entire sample due to the primary $Al_{11}Ce_3$ phase formation caused by low cooling rates. The compressive yield strength of wedge-cast Al-9Ce and Al-12Ce alloys in sections 1 and 5 are compared in Figure 8b. Consistent with the microhardness conclusions, Al-9Ce exhibits reduced strength when slow-cooled (Section 5), whereas Al-12Ce maintains consistent strength, directly controlled by the volume fraction of the eutectic strengthening phase.

The cooling rates in this study cover the typical ranges for industrial permanent die casting and possibly the higher end of sand casting. Our study establishes the correlation between composition, cooling rate, microstructure, and mechanical properties in Al-Ce/RE alloys, which holds significant practical importance. The formation of primary $Al_{11}Ce_3$ phases is found to be detrimental, with minimal strengthening effect but reducing ductility [50], and therefore should be avoided. The cooling rates in casting pieces can vary depending on the mold geometry. The current findings indicate that the material's strength remains unaffected by local cooling rates as long as these rates exceed a certain threshold. Hence, it is crucial to control the cooling rates across the casting pieces above a critical threshold throughout the casting process of near-eutectic Al-Ce alloys to prevent the formation of primary $Al_{11}Ce_3$ phases.

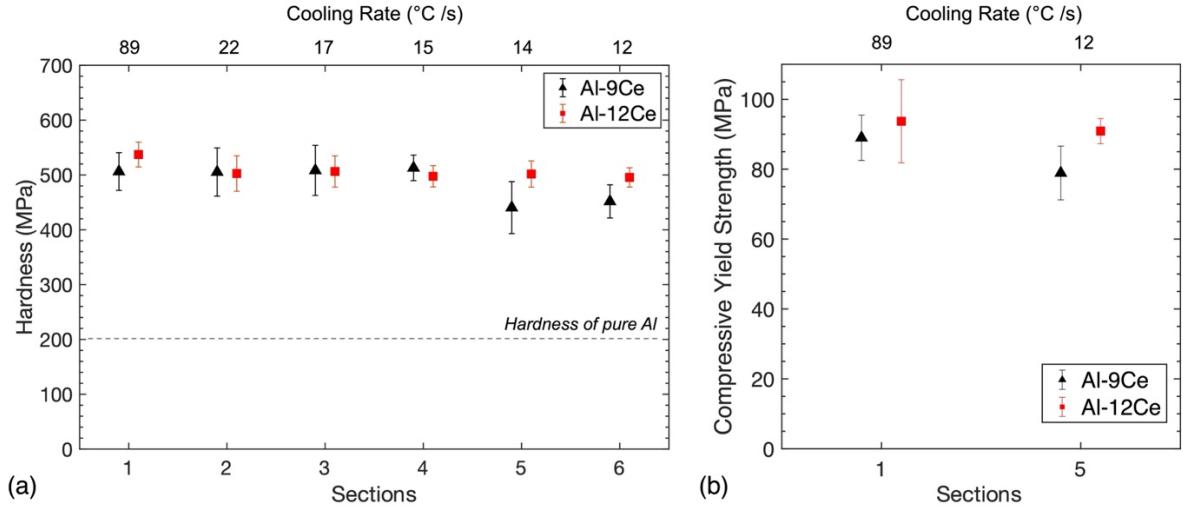

Figure 8. Variation of (a) microhardness and (b) compressive yield strength at ambient temperature for the six sections of the wedge-cast Al-9Ce and Al-12Ce alloys.

### 3.3. Effects of Replacing Ce by Mischmetal: Alloy Properties, Energy and $CO_2$ Reduction

The primary impetus for substituting Ce with MM stems from the aim to reduce energy consumption and $CO_2$ emissions for aluminum alloy production. This calculation is based on the processing stages involved in extracting these elements from bastnaesite ore, as detailed in the flowchart in Figure 9, relevant to the Mountain Pass mine in California, the largest rare-earth mine in the U.S. [22]. The initial stages of mining, extraction, and initial separation are performed on the ore containing a mixture of RE elements. The light RE elements can then be electrolyzed directly to produce MM without prior separation to isolate Ce from Nd and Pr. The production of MM omits the additional separation step required for Ce, resulting in a notable energy saving – a 15% decrease from 73.2 MJ/kg to 62.5 MJ/kg, as detailed in Figure 9.

The estimation of the $CO_2$ footprint for cerium production at Mountain Pass is challenging due to the confidentiality of their separation processes. Alternatively, the $CO_2$ footprint for Ce production is 18.6 $CO_2$-eq kg/kg (as determined in 2018 from Ref. [51]) for the Bayan Obo mine in China, one of the world's largest rare-earth mines. Given that most energy consumption in Ce/MM production is attributed to electricity use, the 15% energy consumption reduction is expected to translate into a similar reduction in $CO_2$ emissions. Consequently, the estimated $CO_2$ footprint for MM production at the Bayan Obo mine is 15.8 $CO_2$-eq kg/kg. The Bayan Obo and Mountain Pass mines produce RE elements from similar bastnaesite ores, using comparable methods. However, due to factors such as the electricity emission rate difference between the U.S. and China, the $CO_2$ footprint for Ce production in the two mines is likely to be different.

In summary, substituting Ce with MM yields a 15% saving in energy (from 73.2 to 62.5 MJ/kg for the Mountain Pass mine) and $CO_2$ emissions (from 18.6 to 15.8 $CO_2$-eq kg/kg for the Bayan Obo mine). Utilizing recycled Ce, La, and/or Nd [23–25] can further enhance these savings by omitting some pre-electrolysis steps (e.g., mining, extraction, separation, transport). Our work also highlights the flexibility of MM compositions in Al-MM-based alloys, enhancing the practicality of MM recycling by eliminating the need for composition adjustment post-recycling. This aspect significantly contributes to the environmental and economic benefits of using MM (or low-purity Ce with fluctuating concentrations of La and other RE), rather than high-purity Ce, in aluminum alloy production.

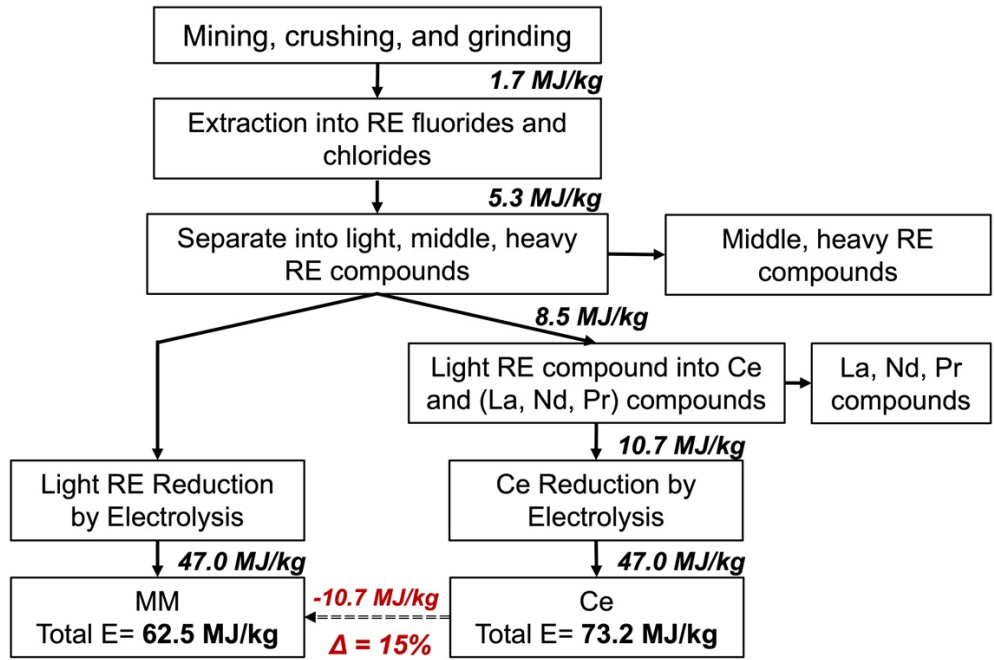

Figure 9. Simplified flow chart for the production process for rare-earth metals from bastnaesite ore, with energy consumption associated with each step relevant to the ore from the Mountain Pass mine (CA, USA). Calculations are given in Supplementary Information, using literature data [22].

## 4. Conclusions

This research studied the effect of replacing Ce with mischmetal (MM) by systematically investigating cast Al-RE with different RE compositions, assessing the microstructure, hardness, tensile properties, creep resistance, and thermal stability under as-cast and thermally exposed conditions. The influence of the casting cooling rate is further studied. The key conclusions are:

1. Binary Al-RE alloys (RE = Ce, CeLa, 2CeLa, or 54Ce27La19Nd) exhibit near-eutectic and hyper-eutectic microstructures for the Al-9RE and Al-12RE compositions (wt%), respectively, but with comparable as-cast hardness (~525 MPa) and coarsening resistance. The additional 3 wt% of RE in Al-12RE forms primary $Al_{11}RE_3$ phases without increasing the volume fraction of $Al_{11}RE_3$ in the eutectic regions, with no improvement in load transfer strengthening and alloy hardness.

2. Al-9MM (MM specifically stands for 54Ce27La19Nd) shows minimal hardness reduction after thermal exposure at 300 or 350 ºC for up to 11 weeks. After thermal exposure at 400 ºC for up to 8 weeks, Al-9/12RE alloys exhibit a modest hardness reduction of ~15%, outperforming eutectic Al-12.6Si and Al-6.4Ni with hardness reduction of ~40 and ~27%, respectively. Microstructural analysis shows minimal phase coarsening for the 300 or 350 ºC/11 weeks conditions, but obvious phase coarsening and spheroidization for the 400 ºC/8 weeks condition.

3. Al-9RE alloys exhibit consistent tensile properties, with yield stress ~ 55 MPa, ultimate tensile strength ~ 130 MPa, and fracture strain ~ 8 %. Mischmetal composition has minimal impact on either microstructure or mechanical properties, highlighting the compositional flexibility of mischmetal in alloy development.

4. When creep-tested at 300 ºC, Al-9MM exhibits an apparent stress exponent n=9 and a threshold stress of 18 MPa. The alloy has higher creep resistance than most Al-Sc-Zr precipitation-strengthened alloys and Al-Mg/Mn solid-solution-strengthened alloys. However, Al-9MM is less creep-resistant than eutectic-strengthened Al-6.4Ni and Al-10Ce-5Ni alloys, consistent with the higher volume fraction and finer fibrous morphology of the eutectic phases present in the latter alloys.

5. The microstructure of wedge-cast Al-9Ce transitions from hypo- to hyper-eutectic as cooling rates decrease from 89 to 17 °C /s, whereas Al-12Ce consistently exhibits hyper-eutectic microstructures. The $Al_{11}Ce_3$ lamellae in both alloys become somewhat finer and more closely spaced with increasing cooling rates. Al-9Ce maintains steady hardness and compressive yield strength at cooling rates between 89 to 15 °C /s but shows reduced hardness and strength at lower rates. Conversely, the hardness of Al-12Ce remains unchanged across all cooling rates.

6. Replacement of Ce by MM from ore yields a 15% saving in energy and $CO_2$ footprint during the mining and refining process. Using recycled Ce- and La-compounds rather than ore can further enhance these savings.

**Data availability** - All data can be accessed by reaching out to the corresponding author.

**Author contributions** - Jie Qi:   Methodology, Investigation, Data acquisition and analysis, Visualization, Writing - original draft, Writing – review & editing. Erin C. Bryan: Investigation, Data acquisition and analysis, Visualization, Writing – review & editing. David C. Dunand: Methodology, Conceptualization, Funding acquisition, Supervision, Writing - review & editing.

**Declaration of competing interest** - DCD discloses a financial interest in NanoAl, LLC (a part of Aluminum Dynamics, LLC and Steel Dynamics, Inc.) which is active in the field of aluminum alloys.

**Acknowledgments** - This research was sponsored by award DE-EE0010221 under the Advanced Materials and Manufacturing Technologies Office, Office of Energy Efficiency and Renewable Energy, the U.S. Department of Energy.  Specimen preparation by arc melting was performed at the Northwestern University Center for Atom Probe Tomography (NUCAPT).  NUCAPT received support from the MRSEC program (NSF DMR-2308691) at the Materials Research Center, the SHyNE Resource (NSF ECCS-2025633), and the Paula M. Trienens Institute for Sustainability and Energy at Northwestern University. This work made use of the EPIC facility of Northwestern University's NUANCE Center, which has received support from the SHyNE Resource (NSF ECCS-2025633), the IIN, and Northwestern's MRSEC program (NSF DMR-1720139).  The authors thank (i) Mr. David Weiss (Eck Industries, WI) and Mr. Scott Rose (Boeing Research & Technology) for numerous useful discussions; (ii) Prof. Xiaohan Du (City University of Hong Kong) for her assistance in FEM modeling; (iii) Ms. Katalin Maji (Evanston Township High School, IL) for her assistance in sample processing and hardness measurements and (iv) Dr. Clement N. Ekaputra (Northwestern University, IL) for his assistance in cooling rate measurements.